# Hole compensation effect in III-Mn-V dilute ferromagnetic semiconductors

Chi Xu[1,3], Mao Wang[1,3], Ye Yuan[1,2,4,*], Gerard Larkin[1], Manfred Helm[1,3] and Shengqiang Zhou[1]

[1]Helmholtz-Zentrum Dresden-Rossendorf, Institute of Ion Beam Physics and Materials Research, Bantzner Landstr. 400, 01328, Dresden, Germany

[2]King Abdullah University of Science and Technology (KAUST), Physical Science and Engineering Division (PSE), Thuwal 23955-6900, Saudi Arabia

[3]Technische Universität Dresden, 01062, Dresden, Germany

[4]Songshan Lake Materials Laboratory, Dongguan, Guangdong, 523808, People's Republic of China

**Abstract**

A systematic study of hole compensation effect on magnetic properties, which is controlled by defect compensation through ion irradiation, in (Ga,Mn)As, (In,Mn)As and (Ga,Mn)P is represented in this work. In all materials, both Curie temperature and magnetization decrease upon increasing the hole compensation, confirming the description of hole mediated ferromagnetism according to the *p-d* Zener model. The material dependence of Curie temperature and magnetization versus hole compensation reveals that the manipulation of magnetic properties in III-Mn-V dilute ferromagnetic semiconductors by ion irradiation is strongly influenced by the energy level location of the produced defect relative to the band edges in semiconductors.

[*] Email: ye.yuan@kaust.edu.sa

## Introduction

Dilute ferromagnetic semiconductors (DFSs), where a fraction of cation site atoms are substituted by $3d$ magnetic ions, exhibit electrically tunable ferromagnetism, thus providing a promising prospect for spintronics [1-3]. Many researchers have focused on III-Mn-V DFSs since Mn acts as the source of both local moments and holes, and the holes mediate ferromagnetism through long-range $p$-$d$ exchanging coupling. To date, many approaches have been employed to flexibly tune magnetic properties via manipulating hole concentration in DFSs, e.g. electrical gating [2, 4-6], low-temperature annealing [7-9], donor/acceptor doping [10, 11] as well as irradiation induced compensation [12, 13]. For electrical gating, the samples have to be as thin as several nano-meters due to the extremely high carrier concentrations [2, 4, 6]. In (Ga,Mn)As co-doped with Be/Zn acceptors, Mn ions are kicked out from substitutional sites to interstitial sites [10, 14]. Note that interstitial Mn ions are double donors in GaAs and can develop anti-ferromagnetic coupling [15, 16].

The high-energy particle irradiation has been used to modify the carrier concentration in various semiconductors [13, 17, 18]. The hydrogen-like defects in the semiconductor matrix produced by ion irradiation drive the Fermi energy towards the same location relative to the vacuum level which is independent of semiconductor materials [19, 20]. As a consequence, due to the different locations of band edges in different semiconductors, irradiation leads to totally different phenomena. For instance, in $In_xGa_{1-x}N$ ion irradiation will induce donor-like defects x>0.34 and acceptor-like defects when x<0.34 [19]. Interestingly, such a method presents an advantage in modifying the properties of DFSs [12, 18, 21, 22]: While keeping Mn atoms at the substitutional sites, the ion beam produces defects in the matrix, further shifting the Fermi energy to a typical location [12, 23, 24]. Therefore, the different locations of band edges in various DFSs [19, 25] result in different manipulation effects on their magnetic properties [12, 18]. Thus, in order to systematically study and compare the contribution of hole concentration in DFS materials, irradiation will be an ideal approach. Moreover, ion irradiation is a well-developed semiconductor-chip technology. It can be easily extended to industry scale applications once being demonstrated in research labs.

In the present work, we carry out a systematic investigation of the hole compensation effect on magnetic properties in (Ga,Mn)As, (In,Mn)As and (Ga,Mn)P. Through the comparison between the Curie temperature ($T_C$) and magnetization vs. irradiation fluences, a full picture of the influence of dopant-like defects on magnetization in III-Mn-V DFSs is illustrated.

## Experiment

The (In,Mn)As and (Ga,Mn)P samples were prepared by implanting Mn ions into (001) InAs and GaP wafers. The wafer normal was tilted by 7° to the ion beam to avoid channeling during ion implantation. A XeCl excimer pulsed laser (Coherent ComPexPRO201, wavelength $\lambda = 308$ nm and pulse duration $\tau = 30$ ns) was used to recrystallize the amorphous as-implanted layers. Different energy densities were used to get the optimal epitaxial structure and the highest Curie temperature for different films, which is 0.20 and 0.40 J/cm$^2$ for (In,Mn)As and (Ga,Mn)P, respectively [26, 27]. The Mn oxide layer formed during the laser annealing process was removed by immersing in a (1:10) HCl solution for 24 hours. After the etching, the accurate values of Mn concentration were determined by secondary ions mass spectrometry (SIMS) using a Cameca IMS 6F microanalyser.

After the preparation, all of the samples were irradiated by He$^+$ ions with various fluences. The irradiation energies were carefully chosen to match the thickness of the Mn implanted layer. The displacement per atom (DPA), which is the number of times that a target atom is displaced during irradiation, is used to quantify the irradiation effect in all samples instead of the irradiation fluences. The DPA is also calculated by SRIM. Corresponding implantation parameters are listed in Table 1. It can be easily observed in the table that DPA is material-dependent and proportional to He irradiation fluence.

Table 1 Sample definitions and related parameters.

| Sample No. | Mn conc. (%) | He irradiation | | |
|---|---|---|---|---|
| | | Energy (keV) | Fluence (cm$^{-2}$) | Peak DPA |
| InAs-0 | 6.1 | 17 | 0 | 0 |
| InAs-2.4e12 | 6.1 | 17 | $2.4\times10^{12}$ | $0.17\times10^{-3}$ |
| InAs-6.0e12 | 6.1 | 17 | $6.0\times10^{12}$ | $0.42\times10^{-3}$ |
| InAs-1.2e13 | 6.1 | 17 | $1.2\times10^{13}$ | $0.84\times10^{-3}$ |
| InAs-1.8e13 | 6.1 | 17 | $1.8\times10^{13}$ | $1.3\times10^{-3}$ |
| InAs-2.4e13 | 6.1 | 17 | $2.4\times10^{13}$ | $1.7\times10^{-3}$ |
| InAs-3.0e13 | 6.1 | 17 | $3.0\times10^{13}$ | $2.1\times10^{-3}$ |
| GaP-0 | 5.5 | 6 | 0 | 0 |
| GaP-5.0e12 | 5.5 | 6 | $5.0\times10^{12}$ | $0.4\times10^{-3}$ |
| GaP-1.0e13 | 5.5 | 6 | $1.0\times10^{13}$ | $0.8\times10^{-3}$ |
| GaP-2.0e13 | 5.5 | 6 | $2.0 \times 10^{13}$ | $1.6 \times 10^{-3}$ |
| GaP-3.0e13 | 5.5 | 6 | $3.0 \times 10^{13}$ | $2.4 \times 10^{-3}$ |

The measurements of magnetic properties were carried out with a SQUID-VSM (Superconducting Quantum Interference Device with Vibrating Sample Magnetometer, Quantum Design, Inc.). For all magnetization hysteresis curves at 5 K, the diamagnetic signal from the substrate was carefully subtracted using the room temperature *M-H* curves where only the diamagnetic signal contributes. For the temperature dependent thermo-remnant magnetization (TRM) measurements, the sample was cooled down from 300 K to 5 K under an external field of 1 kOe, then the field was reset to zero and data were collected during the warming process. It should be noted that all magnetic measurements were carried out along the magnetic easy axis, which is the [001] direction for (In,Mn)As samples and $[1\bar{1}0]$ for (Ga,Mn)P samples.

## Results and discussion

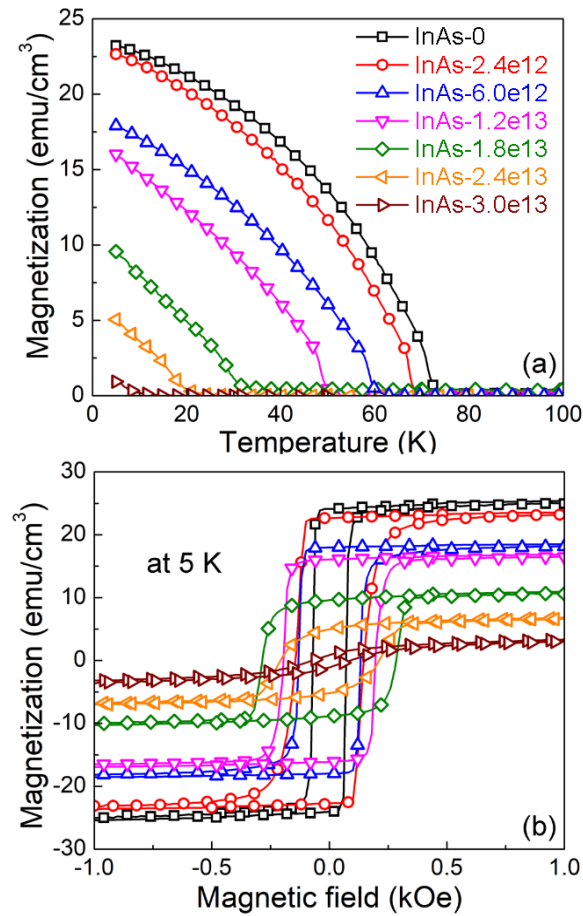

**Figure 1.** Magnetic properties of (In,Mn)As before and after He$^+$ irradiation with various fluences.

(a) Temperature dependent thermo-remnant magnetization; (b) Magnetic-field dependent magnetization at 5 K.

Figure 1 shows the temperature dependent thermo-remnant magnetization (TRM) and the magnetization versus field at 5 K for (In,Mn)As samples before and after irradiation by He ions at different fluences. From figure 1(a), one observes that the Curie temperature ($T_C$) drops upon increasing $He^+$ irradiation fluences through the whole range, from initially 73 K to 11 K when the fluence reaches $3\times10^{13}$ $cm^{-2}$. Moreover, the reduction of $T_C$ is accompanied by a descending remnant magnetization over the whole temperature range, e.g. the TRM at 5 K gradually changes from 23.4 down to 2.1 emu/cm$^3$ with increasing irradiation fluences. As shown in figure 1(a), it is noted that with enhancing irradiation fluences the TRM curves progressively deviate from the concave curvature. This can be explained by the generation of superparamagnetic phases when the hole concentration is not sufficiently high for coupling all Mn magnetic moments [4, 28]. In addition to the $T_C$ and TRM, the saturation magnetization exhibits the same tendency, decreasing with rising irradiation fluences, as shown in figure 1(b), from 26.1 to 3.9 emu/cm$^3$. The coercivity ($H_C$) increases with increasing irradiation fluences, showing a maximum value about 284 Oe, and then decreases when the samples are rather heavily compensated. Similar results in He irradiated (Ga,Mn)As have been reported in our former work [12] where both $T_C$ and magnetization also decrease with increasing irradiation fluences. However, $H_C$ of He irradiated (Ga,Mn)As rises monotonically with increasing irradiation fluence.

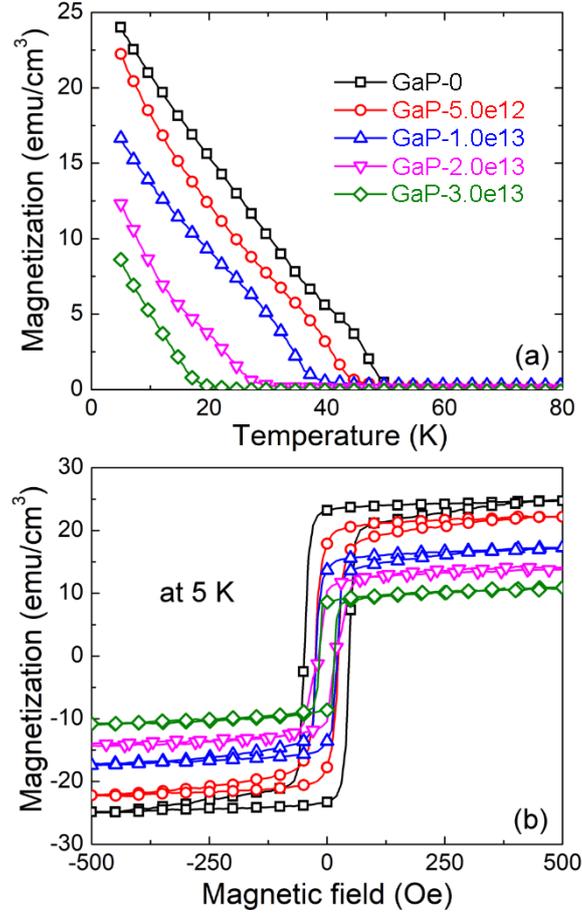

**Figure 2.** Magnetic properties of (Ga,Mn)P before and after He$^+$ irradiation. (a) Thermo-remnant magnetization versus temperature; (b) Magnetization versus magnetic field at 5 K.

GaP exhibits a band gap of 2.26 eV which is larger than the gap of GaAs (1.4 eV) and InAs (0.35 eV), therefore it reveals different features when it is doped with Mn. To date, high-quality epitaxial (Ga,Mn)P thin film can be only obtained by ion implantation combined with pulsed laser melting [27, 29-31]. Unexpectedly, even though $T_C$ approaches 60 K, the sample is still electrically insulating and the Mn impurity stays separated from the GaP valence band [29-31]. Such a phenomenon is explained by the strong *p-d* coupling between localized holes and Mn moments, which enables the super-exchange interaction coupling between the dilute magnetic moments [3]. Nevertheless, the essence of hole-mediated magnetization is not changed in (Ga,Mn)P, therefore the magnetization could be also tuned by compensation induced by ion irradiation as in (Ga,Mn)As and (In,Mn)As. As shown in figure 2, the magnetic properties are tuned similarly as in (Ga,Mn)As and (In,Mn)As: both $T_C$ and magnetization decrease upon increasing fluence. More details can be observed in figure 2(a), both virgin and irradiated samples present temperature dependent TRM curves which is not mean-field like. With increasing He ion fluence from 0 to $2.4×10^3$ cm$^{-2}$, the remnant magnetization and $T_C$ decrease from the initial value 24.0 emu/cm$^3$ to 8.6 emu/cm$^3$ and 50.7 K to 19.6 K, respectively. Due to the compressive strain in the (Ga,Mn)P grown on the GaP substrate, the magnetic easy axis is along the in-plane direction, as shown in figure 2(b) that the hysteresis loops are highly square like when the magnetic field is applied along the in-plane direction. For the changes in hysteresis loops, all samples show similar results as in (Ga,Mn)As and (In,Mn)As: both saturation and remanent magnetization decrease upon increasing ion irradiation fluence. Additionally, $H_C$ firstly shows a reduction behavior, and finally becomes constant.

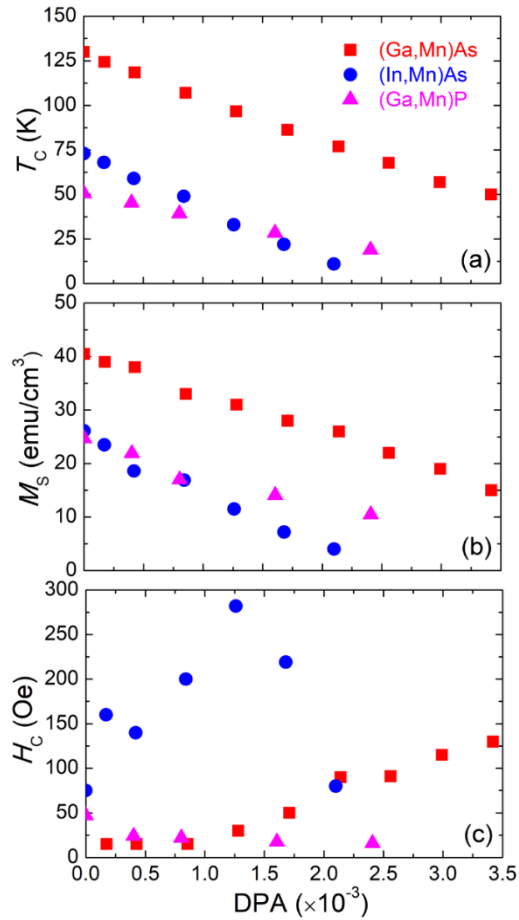

**Figure 3.** Curie temperature ($T_C$) (a), saturation magnetization ($M_S$) (b) as well as coercive field ($H_C$) for the magnetic easy axis (c) at 5 K versus DPA for (Ga,Mn)As (squares), (In,Mn)As (circles) and (Ga,Mn)P (triangles). Data of (Ga,Mn)As are taken from our former work [12].

To compare the hole-compensation effect on the magnetic properties in (Ga,Mn)As, (In,Mn)As and (Ga,Mn)P, the Curie temperature, saturation magnetization, and coercivity versus DPA are plotted together in figure 3. All of these three materials show a linear dependence of $T_C$ versus DPA, where the slope of (In,Mn)As is the largest of three samples while the one of (Ga,Mn)P is the smallest. The saturation magnetization shows a similar characteristic as $T_C$ vs. DPA. The results indicate that the defect induced compensation influences differently the magnetic properties of (Ga,Mn)As, (In,Mn)As and (Ga,Mn)P, and more detailed analysis will be carried out in the later discussion. Interestingly, $H_C$ exhibits totally different dependence when DPA raises: it rises and decreases monotonically for (Ga,Mn)As and (Ga,Mn)P, respectively, while for (In,Mn)As it shows a maximum value of around 300 Oe. However, it is worth to mention that a switching of the magnetic anisotropy does not appear here [32, 33].

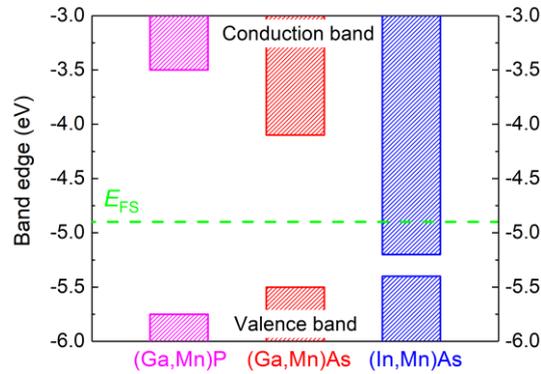

**Figure 4.** Position of the valence band maxima, the conduction band minima and the Fermi-level stabilization energy (dash line) for (Ga,Mn)P, (Ga,Mn)As, and (In,Mn)As relative to the vacuum level [20].

From our previous work on (Ga,Mn)As, there is no detectable increase of interstitial Mn after He ion irradiation [10]. Thus, the reduced $M_S$ and $T_C$ are due to a decreased hole concentration. From figures 3(a) and 3(b), it is concluded that the irradiation induced defects affect the three DFSs differently, and the comparison of $M_S$ and $T_C$ indicates that (In,Mn)As is most sensitive to irradiation induced defects, (Ga,Mn)P is the least sensitive and (Ga,Mn)As is in between. The difference can be understood using the amphoteric defect model [23, 24]. Figure 4 shows the valence band maxima (VBM) and the conduction band minima (CBM), relative to the vacuum level in GaAs, InAs and GaP samples [24]. The dashed line in the figure stands for the Fermi-level stabilization energy $E_{FS}$, which is the universal energy level of the irradiation-induced point defects. This model describes the amphoteric formation of defects: acceptor (donor) defects are formed when the Fermi level $E_F > E_{FS}$ ($E_F < E_{FS}$), resulting in a shift of $E_F$ towards $E_{FS}$ and finally reaches $E_{FS}$ as a saturation value at a high enough irradiation fluence. The location of $E_{FS}$ does not rely on the doping type or the Fermi level and is always approximately at a constant value of 4.9 eV below the vacuum level. Therefore, according to this model, the discussion can be simplified, as for the formation of defects, the only difference among these samples is the location of $E_{FS}$ relative to the band edges. In the case of InAs, $E_{FS}$ stays in the conduction band, which indicates that the irradiation induced defects will generate itinerant electrons in the InAs matrix till driving the Fermi energy into the conduction band. On the contrary, the $E_{FS}$ in GaAs and GaP is deeply in the band gap, which is 0.8 and 1.4 eV below the bottom of the conduction band of GaP and GaAs, respectively, suggesting that the Fermi level would be upshifted from the valence band till deeply in the band gap. Such a deeply located $E_{FS}$ indicates that the hole compensation in these two semiconductors is not as intensive as in InAs, which explains the smaller influence of defects on hole concentration, further on $T_C$ and magnetism. For the comparison between (Ga,Mn)As and (Ga,Mn)P, the intensity of hole compensation can be evaluated by the gap between $E_{FS}$ and the minima of the conduction band, where the smaller band gap in GaAs results in more intensive compensation, thus leading to the larger slope of d$T_C$/dDPA (d$M$/dDPA) when compared to GaP. Furthermore, the ferromagnetism in (Ga,Mn)P is mediated by the localized holes in the impurity band [21]. This may also explain the insensitivity of (Ga,Mn)P subjected to ion irradiation.

## Conclusions

In this work, the effect of irradiation induced hole compensation on magnetism in III-Mn-V DFSs was systematically investigated in (Ga,Mn)As, (In,Mn)As and (Ga,Mn)P irradiated by He$^+$ with different fluences. Steady decrease of the Curie temperature and saturation magnetization was observed upon increasing irradiation fluences. The hole compensation effect is the strongest in (In,Mn)As while weakest in (Ga,Mn)P compared with (Ga,Mn)As. The amphoteric defect model can be used to understand such different hole-compensation effect resulted from ion irradiation.

## Acknowledgements

Support by the Ion Beam Center (IBC) at HZDR is gratefully acknowledged. The authors thank Prof. B. L. Gallagher, Dr. K. W. Edmonds, Dr. R. P. Campion and Dr. A. W. Rushforth (University of Nottingham) for providing (Ga,Mn)As samples. The support from Rafal Jakiela (Polish Academy of Sciences) for SIMS measurement is acknowledged. This work is funded by the Helmholtz-Gemeinschaft Deutscher Forschungszentren (HGF-VH-NG-713). The author C. Xu thanks financial support by Chinese Scholarship Council (File No. 201506680062).